\documentclass[twocolumn,aps,9pt,showpacs,showkeys,]{revtex4}
\usepackage[dvips]{graphicx}

\begin{document}
\title{The atomic-start description of NiO$^\spadesuit$}
\author{R. J. Radwanski}
\affiliation{Center of Solid State Physics, S$^{nt}$Filip 5, 31-150 Krakow, Poland,\\
Institute of Physics, Pedagogical University, 30-084 Krakow,
Poland}
\author{Z. Ropka}
\affiliation{Center of Solid State Physics, S$^{nt}$Filip 5,
31-150 Krakow, Poland}
\homepage{http://www.css-physics.edu.pl}
\email{sfradwan@cyf-kr.edu.pl}

\begin{abstract}
We have calculated magnetic properties and the electronic
structure of NiO both in the paramagnetic and in
magnetically-ordered state as well as zero-temperature properties
and thermodynamics within the strongly-correlated crystal-field
approach. It is in agreement with a Mott's suggestion that NiO is
an insulator due to strong electron correlations. We have
quantified crystal-field, spin-orbit and magnetic interactions of
the Ni$^{2+}$ ion in NiO. We have obtained that E$_{dd}$ $\gg$
E$_{CF}$(=2.0 eV)$\gg$E$_{s-o}$(=0.29 eV)$\gg$E$_{mag}$(=0.07 eV).
The orbital moment of 0.54 $\mu_{B}$ amounts at 0 K, in the
magnetically-ordered state, to about 20\% of the total moment
(2.53 $\mu_{B}$). Our studies indicate that it is the highest time
to "unquench" orbital magnetic moment in 3d solid-state physics
and the necessity to take always into account strong intra-atomic
correlations among d electrons and the intra-atomic spin-orbit
coupling.

\pacs{75.25.+z, 75.10.Dg} \keywords{Crystalline Electric Field, 3d
oxides, magnetism, spin-orbit coupling, NiO}
\end{abstract}
\maketitle

NiO attracts large attention of the magnetic community by more
than 50 years. Despite of its simplicity (two atoms, NaCl
structure, well-defined antiferromagnetism (AF) with T$_{N}$ of
525 K) and enormous theoretical and experimental works the
consistent description of its properties, reconciling
its insulating state with the unfilled 3$d$ band is still not reached \cite%
{1,2,3,4,5,6,7,8,9,10}.

The aim of this paper is to report the calculations of the
magnetic moment, specific heat and the low-energy electronic
structure of bulk NiO within the conventional ionic picture
completed with inter-site spin-dependent interactions. We
attribute this moment and this low-energy electronic structure to
the Ni$^{2+}$ ions in the rhombohedrally distorted NaCl structure.
The approach used can be called the quasi-atomic approach as the
starting point for the description of a solid is the consideration
of the atomic-like electronic structure of the constituting
atoms/ions. In the present case we start from the atomic-like
electronic structure of eight electrons in the 3d shell of the
Ni$^{2+}$ ions in the oxygen octahedron. A macroscopic body of NiO
with the NaCl structure is built up from such face-sharing
octahedra.

We have treated the 8 outer electrons of the Ni$^{2+}$ ion as
forming the strongly-correlated atomic-like electron system
3d$^{8}$. Its ground term is described by
two Hund's rules yielding $S$=1 and $L$=3, i.e. the ground term $^{3}F$ \cite%
{11}. In a solid such the localized strongly-correlated electron
system interacts with the charge and spin surroundings. The charge
surrounding has predominantly the octahedral symmetry owing to the
NaCl-type of structure of NiO. Our Hamiltonian for NiO consists of
two terms: the single-ion-like term $H_{d}$ of the 3$d^{8}$ system
and the $H_{d-d}$ intersite spin-dependent term. Calculations
somehow resemble those performed for rare-earth systems, see e.g.
Ref. \cite{12} and they have been recently applied successfully to
3$d$ compounds \cite{10,13}. For the calculations of electronic
states of the quasi-atomic single-ion-like Hamiltonian of the
3$d^{8}$ system we take into account the crystal-field
interactions of the octahedral symmetry and the spin-orbit
coupling (the octahedral
CEF parameter $B_{4}$=+21 K (=+1.81 meV), the spin-orbit coupling constant $%
\lambda _{s-o}$ = -480 K (= -41 meV) \cite{11}). It is very
important to know, that the positive sign of $B_{4}$ comes
directly from {\it ab initio} calculations for the oxygen anion
octahedral surroundings. We have chosen the value of B$_{4}$ of 21
K in order to reproduce experimentally observed an absorption peak
at 1.06-1.13 eV \cite{14}, see also Fig. 5 of Ref. \cite{8}. The
single-ion states under the octahedral crystal field and the
spin-orbit coupling have been calculated by consideration of the
Hamiltonian \cite{15}:

\begin{figure}[t]
\begin{center}
\includegraphics[width = 5.0 cm]{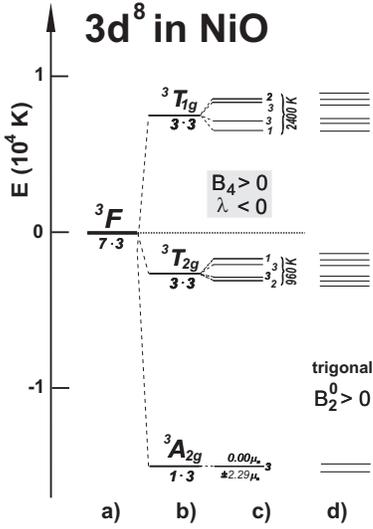}
\end{center} \vspace {-0.3 cm}
\caption{The calculated fine electronic structure of the highly-correlated 3$%
d^{8}$ electronic system in the paramagnetic state. a) the 21-fold
degenerated $^{3}F$ term given by two Hund's rules: $S$=1 and
$L$=3. b) the effect of the cubic octahedral crystal-field with
$B_{4}$=+21 K. c) the combined action of the octahedral crystal
field and the spin-orbit coupling with $\lambda _{s-o}$= -480 K.
d) the effect of a trigonal distortion - the splitting of the
ground triplet amounts to 12 K for B$_{2}^{0}$= +50K. The positive
B$_{2}^{0}$ yields the singlet lower - such a situation is favored
by the Jahn-Teller theorem.}
\end{figure}
\vspace {-0.3 cm}
\begin{equation}
H_{d}=B_{4}(O_{4}^{0}+5O_{4}^{4})+\lambda _{s-o}L\cdot S
\end{equation}
The charge-formed fine electronic structure, shown in Fig. 1,
contains three groups of localized states. The higher groups are
at 1.06-1.14 eV and 1.91-2.05 eV. For low- and room-temperature
properties the lowest triplet, originating from the cubic subterm
$^{3}A_{2g}$, is the most important as the higher states are not
thermally populated. The existence of this triplet in the
spin-orbital space indicates that the Ni$^{2+}$ ion is unstable to
lattice off-octahedral distortions, that splits the triplet. The
value of B$_{4}$= +21 K corresponds to another characteristics of
the cubic crystal field Dq of 1260 K (=875 cm$^{-1}$).
Historically for a configuration with a F ground term (d$^{2}$,
d$^{3}$, d$^{7}$, d$^{8}$) the energy separation between
$^{3}$A$_{2g}$ and $^{3}$T$_{2g}$ subterms is denoted as 10Dq. It
should be noted that in theoretical approaches different values
for 10Dq are considered like only 0.5 eV in Ref. \cite{2,5}. For
getting the proper insulating gap an enormous splitting between
spin-up and spin-down e$_{g}$ states of 10.8 eV was introduced in
Ref. \cite{4}, that according to us is not physically justified.

For the description of the trigonal distortion observed in NiO the
cubic Hamiltonian (1) is necessary to transform in order to have
the z quantization axis along the cube diagonal. In case of the
cubic NaCl structure the trigonal distortion is realized by
stretching or compressing along the cube diagonal. The CEF
Hamiltonian to describe such the distortion takes the form:
\vspace {-0.1cm}
\begin{equation}
H_{d}=-2/3 B_{4}(O_{4}^{0}-20\sqrt{2}O_{4}^{3})+ B_{2} O_{2}^{0}
\end{equation}
The obtained eigenfunctions of the lowest spin-orbital triplet,
for $B_{4}$=+21 K and $\lambda _{s-o}$= -480 K yielding $\lambda
_{s-o}$/10Dq=0.038, take a form (only
$|$$L$$_{z}$,$S$$_{z}$$\rangle$ components of the full
$|$$L,S,L$$_{z}$,$S$$_{z}$$\rangle$ function are shown):

$\psi$$_{1}$ = 0.740 $|$0,+1$\rangle$ -0.522$|$+3,+1$\rangle$ +
0.418$|$-3,+1$\rangle$

$\psi$$_{0}$ = 0.744 $|$0,0$\rangle$ -0.467$|$+3,0$\rangle$ +
0.467 $|$-3,0$\rangle$

$\psi$$_{-1}$ = 0.740 $|$0,-1$\rangle$ +0.522$|$-3,+1$\rangle$ -
0.418$|$+3,+1$\rangle$

For these functions $S_{z}$=$\pm$0.9945 and $L_{z}$=$\pm$0.297,
what give the total moment of $m_{z}$=$\pm$2.2883 $\mu_{B}$. The
$L_{z}$ value of 0.297 is close to the first-order spin-orbit
correction (Ref. \cite{11}, p. 450) of
8$\cdot$$\lambda_{so}$/10Dq=8$\cdot$0.038=0.30.

For the trigonal distortion for $B_{2}^{0}$ $>$0 the triplet
splits into lower singlet and higher doublet. For $B_{2}^{0}$ $<$0
the doublet is lower. For $B_{2}^{0}$=+50 K the splitting of the
ground triplet amounts to 12 K.

The magnetic field, external or internal in the case of the
magnetically-ordered state, polarizes two states of the doublet,
as is shown in Fig. 2. The intersite spin-dependent interactions
$H_{d-d}$ cause the (antiferro-)magnetic ordering. They have been
considered in the mean-field approximation with the
molecular-field coefficient {\it n} acting between magnetic
moments $m_{d}$=-($L$+$g_{e}$$S$) $\mu _{B}$, $g_{e}$ = 2.002324.
The value of $n$ in the Hamiltonian \cite{10,13}
\begin{equation} H_{d-d}=n\left( -m_{d}\cdot
m_{d}+\frac{1}{2}\left\langle m_{d}^{2}\right\rangle \right)
\end{equation}
has been adjusted in order to reproduce the
experimentally-observed Neel temperature. The fitted value of $n$
has been found to be -11.55 meV/ $\mu _{B}^{2}$ (= -200 T/$\mu
_{B}$). It means that the Ni ion in the magnetically-ordered state
experiences the molecular field of 503 T (at 0 K). This field as
well as the magnetic moment become smaller with the increasing
temperature and vanish above $T_{N}$.
\begin{figure}[t]
\begin{center}
\includegraphics[width = 5.3 cm]{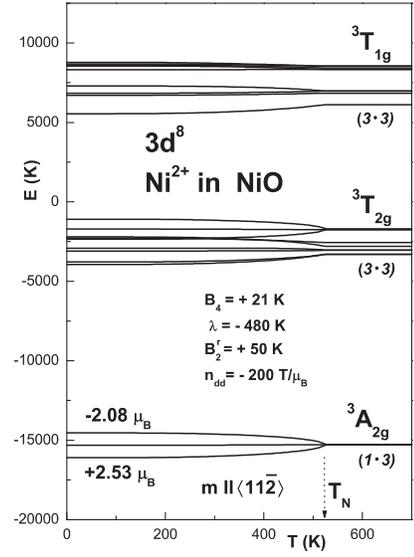}
\end{center} \vspace {-0.5cm}
\caption{The calculated temperature dependence of the fine
electronic structure of the highly-correlated 3$ d^{8}$ electron
system in the magnetically-ordered state below T$_{N}$ of 525 K.
In the paramagnetic state the electronic structure is temperature
independent unless we do not consider, for instance, a changing of
the CEF parameter due to the thermal lattice expansion. }
\end{figure}
The calculated value of the magnetic moment at 0 K in the
magnetically-ordered state amounts to 2.53 $\mu _{B}$, Fig. 2. It
is built up from the spin moment $m_{s}$ of 1.99 $\mu _{B}$
($S$$_{z}$=0.995) and the orbital moment $m_{o}$ of 0.54 $\mu
_{B}$, Fig. 3. The increase of $m_{o}$ in comparison to the
paramagnetic state, $\pm $ 0.26 $\mu _{B}$, is caused by the
further polarization of the ground-state eigenfunction by the
internal molecular magnetic field. The orbital moment is quite
substantial being about 20\% of the total moment. The calculated
by us moment at 300 K amounts to 2.26 $\mu _{B}$ ($m_{s}$= 1.78
$\mu _{B}$, $m_{o}$= 0.48 $\mu _{B}$ ) nicely reproducing the
experimental
result of 2.2$\pm $0.3 $\mu _{B}$ for the Ni moment at 300 K %
\cite{16}. It should be noted that this novel moment value is
remarkably larger than old results of 1.64-1.70 $\mu _{B}$, which
have been generally up to now used in theoretical calculations
\cite{4,6}. The magnetic X-ray experiment of Ref. \cite{16} has
revealed a substantial orbital moment of 0.32 $\mu _{B}$ and the
spin moment of 1.90 $\mu _{B}$ at 300 K.

\begin{figure}[t]
\begin{center}
\includegraphics[width = 6.0 cm]{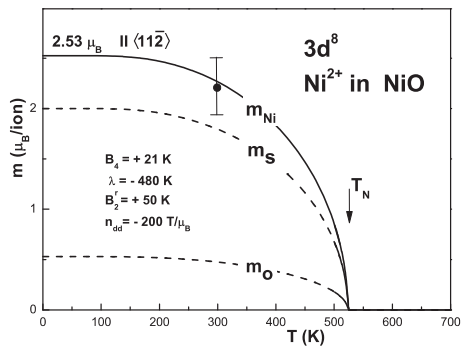}
\end{center}\vspace {-0.5 cm}
\caption{The calculated temperature dependence of the
Ni$^{2+}$-ion moment in NiO. At 0 K the total moment m$_{Ni}$ of
2.53 $\mu _{B}$ is built up from the orbital m$_{o}$ and spin
m$_{s}$ moment of 0.54 and 1.99 $\mu _{B}$, respectively.}
\end{figure}
The trigonal distortion is important for the detailed formation of
the AF structure and the direction of the magnetic moment but it
only slightly influences the spin and orbital momenta values. The
trigonal distortion yielding the singlet charge-formed ground
state leads to the moment direction lying in the plane
perpendicular to the cube diagonal. Exactly such the moment
direction is observed in NiO. Actually, the magnetic ordering
occurs along the $\langle$$11\bar{2}$$\rangle$ direction within
this diagonal perpendicular plane due to a further slight
distortion within the (111) plane \cite{3}.

The present model allows, apart of the ordered moment and its spin
and orbital components, to calculate many physically important
properties like temperature dependence of the magnetic
susceptibility, temperature dependence of the heat capacity (shown
in Fig. 4), the spectroscopic $g$ factor, the fine electronic
structure in the energy window below 3 eV with at least 20
localized states, Fig. 1. The spike-like peak in c(T) at $T_{N}$
is in perfect agreement with experimental data \cite{17} obtained
on a single-crystal specimen that yields ''very large, very narrow
peak of 65 cal/Kmol'' \cite{18}. Ref. \cite{18} provides a
detailed critical analysis of literature experimental data with
the tabulated recommended data, that are shown in Fig. 4. As the
best description of the c(T) dependence in the range 0$<$T$<$250 K
they give the Debye value of $\theta$ of 580 K. In Fig. 4 we
provide our rough summation of the calculated c$_{d}$(T)
contribution and the lattice contribution with a value of $\theta$
of 650 K. We find the overall description of c(T) as remarkably
good owing too so large temperature interval and the
$\lambda$-type peak at T$_{N}$.
\begin{figure}[t]
\begin{center}
\includegraphics[width = 7.2 cm]{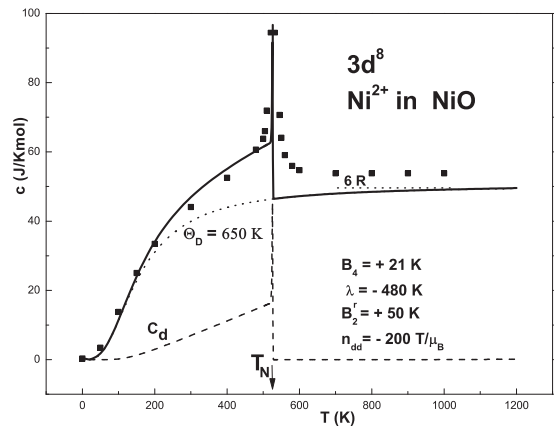}
\end{center}\vspace {-0.45cm}
\caption{The calculated temperature dependence of the 3$d$
contribution $c_{d}(T)$ to the heat capacity of NiO. The spike
peaks to 214 J/Kmol. A value of 6 R, R is a gas constant, for the
high temperature Dulong-Petit heat is shown. Points are the
recommended data by Ref. \cite{18} after a critical analysis of
literature experimental data. }
\end{figure}
The magnetically-ordered state of NiO has lower energy than the
paramagnetic one by 3.4 kJ/mol (= 35 meV/ion) at 0 K, Fig. 2.
Although the molecular field acting on the Ni moment is so large
as 503 T it splits CEF states, however, without their
reorganization in the lowest triplet. The splitting of the spin-up
and spin-down states amounts to 1560 K (135 meV). We think that
this energy excitation among the lowest triplet has been revealed
by Newman and Chrenko almost 50 years ago \cite{14}, who has found
a 240 meV excitation. From temperature dependence they have
concluded that it is connected with the antiferromagnetic
ordering, exactly as comes out from our approach. In our picture
this excitation decreases with increasing temperature vanishing at
T$_{N}$, see Fig. 2.

We would like to point out that our approach should not
be considered as the treatment of an isolated ion only - we consider the Ni$%
^{2+}$ ion in the oxygen octahedron. The physical relevance of our
calculations to macroscopic NiO is obvious - the NaCl structure is
built up from the face sharing Ni$^{2+}$ octahedra. In the perfect
structure all Ni ions experience the same charge (= crystal field)
and spin interactions, what means that all Ni ions have the same
electronic structure. The good reproduction of macroscopic
properties proves that all atoms equally contribute to these
properties. But one should remember that due to a finite size of
each sample always some atoms on the surface, say 1-2 $\%$, will
have another electronic structure like it was discussed in Ref.
\cite{9}.

In contrary to a general conviction within the magnetic community
about the adequateness of the strong crystal-limit to 3d-ion
compounds we have found a remarkably good description of NiO in a
rather weakly-intermediate CEF limit. The octahedral crystal field
is 25 times stronger than the spin-orbit coupling but it does not
break intra-atomic correlations among 3d electrons (after giving
up two electrons during the formation of a solid NiO, from the 4s
states). Such a physical situation is the basis for a developed
Quantum Atomistic Solid State Theory (QUASST) for 3d-atom
containing compounds \cite{19}. We treat our QUASST approach as a
continuation of studies of Van Vleck on correlation of macroscopic
properties with discrete crystal-field states of transition-metal
ions \cite{20}.

In conclusion, we have calculated magnetic properties and the
electronic structure of NiO both in the paramagnetic and in
magnetically-ordered state as well as zero-temperature properties
and thermodynamics. We have quantified crystal-field (the leading
parameter B$_{4}$ = +21 K), spin-orbit (-480 K, i.e. like in the
free ion \cite{11}) and magnetic interactions (B$_{mol}$ of 503 T
and n= -200 T/$\mu_{B}$). In our approach E$_{dd}$ $\gg$
E$_{CF}$(=2.0 eV)$\gg$E$_{s-o}$(=0.29 eV)$\gg$E$_{mag}$(=0.07 eV).
The orbital and spin moment of the Ni$^{2+}$ ion in NiO has been
calculated within the quasi-atomic approach. The orbital moment of
0.54 $\mu _{B}$ amounts at 0 K in the magnetically-ordered state,
to about 20\% of the total moment (2.53 $\mu _{B}$). Despite of
using the full atomic orbital quantum number $L$=3 and $S$=1, the
calculated effective moment from the temperature dependence of the
susceptibility amounts to 3.5-3.8 $\mu _{B}$, i.e. only 20 $\%$
larger value than a spin-only value of 2.83 $\mu _{B}$. Our
success is related to the fact, that we take into account the
existence of very strong correlations among electrons. Good
description of many physical properties indicates that 3$d$
electrons in NiO are in the extremely strongly-correlated limit,
i.e. in the ionic limit confirming {\it a priori} our atomic-start
assumption. In our approach we can trace the Jahn-Teller effect,
in the spin-orbital space, and the breaking of the time reversal
symmetry, equivalent to the appearance of the ordered magnetic
state, at the atomic scale.

We can mention that within the same approach we have described a
few 3d compounds like FeBr$_{2}$ \cite{13}, LaCoO$_{3}$,
Na$_{2}$V$_{3}$O$_{7}$ and recently CoO \cite{21}. The localized
crystal-field states discussed in the present paper are already in
a technical use in a ultrafast manipulation of the
antiferromagnetism of NiO \cite{22}.

$^\spadesuit$ dedicated to John Van Vleck and Hans Bethe, pioneers
of the crystal-field theory, to the 75$^{th}$ anniversary of the
crystal-field theory, and to the Pope John Paul II, a man of
freedom in life and in Science.

\end{document}